\documentclass[prc,aps,twocolumn]{revtex4}
\usepackage{graphicx}
\usepackage{dcolumn}
\usepackage{bm}

\def\tend{\mathop{\to}}

\def\lim{\mathop{\rm {lim}}}

\begin{document}
\draft \preprint{HEP/123-qed}
\title{Renormalization of equations governing nucleon
dynamics \\
and nonlocality in time of the NN interaction. }
\author{Renat Kh.Gainutdinov and Aigul A.Mutygullina}
\address{
Department of Physics, Kazan State University, 18 Kremlevskaya St,
Kazan 420008, Russia } \email{Renat.Gainutdinov@ksu.ru}
\date{\today}

\begin{abstract}
We discuss the problem of renormalization of dynamical equations
which arises in an effective field theory description of nuclear
forces. By using a toy model of the separable NN potential leading
to logarithmic singularities in the Born series, we show that
renormalization gives rise to nucleon dynamics which is governed
by a generalized dynamical equation with a nonlocal-in-time
interaction operator. We show that this dynamical equation can
open new possibilities for applying the EFT approach to the
description of low-energy nucleon dynamics.
\end{abstract}
\maketitle
\narrowtext

\section{Introduction}
\label{sec:level1}

 Understanding how
nuclear forces emerge from the fundamental theory of quantum
chromodynamics (QCD) is one of the most important problem of
quantum physics. To study hadron dynamics at scales where QCD is
strongly coupled, it is useful to employ effective field theories
(EFT's) [1] being an invaluable tool for computing physical
quantities in the theories with disparate energy scales. In order
to describe low energy processes involving nucleons and pions, all
possible interaction operators consistent with the symmetries of
QCD are included in an effective Lagrangian of an EFT. However
such a Lagrangian leads to ultraviolet (UV) divergences that must
be regulated and a renormalization scheme defined. A fundamental
difficulty in an EFT description of nuclear forces is that they
are nonperturbative, so that an infinite series of Feynman
diagrams must be summed. Summing the relevant diagrams is
equivalent to solving a Schr{\"o}dinger equation. However, an EFT
yields graphs which are divergent, and gives rise to a singular
Schr{\"o}dinger potential. For this reason N-nucleon potentials
are regulated and renormalized couplings are defined [2].
Nevertheless, a renormalization procedure does not lead to the
potentials satisfying the requirements of ordinary quantum
mechanics, and consequently after renormalization nucleon dynamics
is not governed by the Schr{\"o}dinger equation. This raises the
question of what kind of equation governs nucleon dynamics at low
energies. The Schr{\"o}dinger equation is local in time, and the
interaction Hamiltonian describes an instantaneous interaction.
This is the main cause of infinities in the Hamiltonian formalism.
In Ref.[3] it has been shown that the use of the Feynman approach
to quantum theory [4] in combination with the canonical approach
allows one to extend quantum dynamics to describe the evolution of
a system whose dynamics is generated by nonlocal in time
interaction.  A generalized quantum dynamics (GQD) developed in
this way has been shown to open new possibilities to resolve the
problem of the UV divergences in quantum field theory [3]. An
equation of motion has been derived as the most general dynamical
equation consistent with the current concepts of quantum theory.
Being equivalent to the Schr{\"o}dinger equation in the particular
case where interaction is instantaneous, this equation permits the
generalization to the case where the interaction operator is
nonlocal in time. Note that there is one-to-one correspondence
between nonlocality of interaction and the UV behavior of the
matrix elements of the evolution operator as a function of
momenta: The interaction operator can be nonlocal in time only in
the case where this behavior is "bad", i.e. in a local theory it
leads to the UV divergences. For this reason one can expect the
nucleon dynamics that follows from renormalization of an EFT to be
governed by the generalized dynamical equation with
nonlocal-in-time interaction operator. In the present paper we
investigate the problem of renormalization of dynamical equations
which arises in the EFT approach. In Sec.II we review the
principal features of the GQD. In Sec.III, by using  a toy model
of the separable NN potential leading to logarithmic singularities
in the Born series, we show that renormalization gives rise to
nucleon dynamics which is governed by the generalized dynamical
equation with a nonlocal-in-time interaction operator. The
dynamical situation that arises in a quantum system of nucleons
after renormalization is investigated in Sec.IV. We show that the
T matrix obtained in Ref.[5] by dimensional regularization of this
model does not satisfy the Lippmann-Schwinger (LS) equation but
satisfies the generalized dynamical equation with a
nonlocal-in-time interaction operator. Finally, in Sec.V we
present some concluding remarks.

\section{Generalized quantum dynamics}
\label{sec:level2} As has been shown in Ref.[3], the
Schr{\"o}dinger equation is not the most general dynamical
equation consistent with the current concepts of quantum theory.
Let us consider these concepts. As is well known, the canonical
formalism is founded on the following assumptions:

(i) The physical state of a system is represented by a vector
(properly by a ray) of a Hilbert space.

(ii) An observable A is represented by a Hermitian hypermaximal
operator $\alpha$. The eigenvalues $a_r$ of $\alpha$ give the
possible values of A. An eigenvector $|\varphi_r^{(s)}>$
corresponding to the eigenvalue $a_r$ represents a state in which
A has the value $a_r$. If the system is in the state $|\psi>,$ the
probability $P_r$ of finding the value $a_r$ for A, when a
measurement is performed, is given by
$$P_{r} = <\psi|P_{V_{r}} |\psi>= \sum_s |<\varphi_r^{(s)}|\psi>|^2, $$
where $P_{V_{r}}$ is the projection operator on the eigenmanifold
$V_r$ corresponding to $a_r,$ and the sum $\Sigma_s$ is taken over
a complete orthonormal set ${|\varphi_r^{(s)}>}$ (s=1,2,...) of
$V_r.$ The state of the system immediately after the observation
is described by the vector $P_{V_{r}}|\psi>.$

In the canonical formalism these postulates are used together with
the assumption that the time evolution of a state vector is
governed by the Schr{\"o}dinger equation. However, in QFT the
Schr{\"o}dinger equation is only of formal importance because of
the UV divergences. Note in this connection that in the Feynman
approach to quantum theory this equation is not used as a
fundamental dynamical equation. As is well known, the main
postulate on which this approach is founded, is as follows [4]:

(iii) The probability of an event is the absolute square of a
complex number called the probability amplitude. The joint
probability amplitude of a time-ordered sequence of events is
product of the separate probability amplitudes of each of these
events. The probability amplitude of an event which can happen in
several different ways is a sum of the probability amplitudes for
each of these ways.

The statements of the assumption (iii) express the well-known law
for the quantum-mechanical probabilities. Within the canonical
formalism this law is derived as one of the consequences of the
theory. However, in the Feynman formulation of quantum theory this
law is used as the main postulate of the theory. The Feynman
formulation also contains, as its essential idea, the concept of a
probability amplitude associated with a completely specified
motion or path in space-time. From the assumption (iii) it then
follows that the probability amplitude of any event is a sum of
the probabilities that a particle has a completely specified path
in space-time. The contribution from a single path is postulated
to be an exponential whose (imaginary) phase is the classical
action (in units of $\hbar$) for the path in question. The above
constitutes the contents of the second postulate of the Feynman
approach to quantum theory. This postulate is not so fundamental
as the assumption (iii), which directly follows from the analysis
of the phenomenon of quantum interference. In Ref.[3] it has been
shown that the first postulate of the Feynman approach (the
assumptions (iii)) can be used in combination with the main
fundamental postulates of the canonical formalism (the assumptions
(i) and (ii)) without resorting to the second Feynman postulate
and the assumption that the dynamics of a quantum system is
governed by the Schr{\"o}dinger equation. As has been shown, such
a use of the main assumptions of quantum theory leads to a more
general dynamical equation than the Schr{\"o}dinger equation.

In the general case the time evolution of a quantum system is
described by the evolution equation
$$|\Psi(t)>=U(t,t_0)|\Psi(t_0)>,$$ where $U(t,t_0)$ is the unitary
evolution operator
\begin{equation}
U^{+}(t,t_0) U(t,t_0) = U(t,t_0) U^{+}(t,t_0) = {\bf
1},
\label{unitary}
\end{equation}
with the group property
\begin{equation}
U(t,t') U(t',t_0) = U(t,t_0), \quad U(t_0,t_0) ={\bf 1}.
\label{compos}
\end{equation}
Here we use the interaction picture. According to the assumption
(iii), the probability amplitude of an event which can happen in
several different ways is a sum of contributions from each
alternative way. In particular, the amplitude
 $<\psi_2| U(t,t_0)|\psi_1>$ can be represented as a sum
of contributions from all alternative ways of realization of the
corresponding evolution process. Dividing these alternatives in
different classes, we can then analyze such a probability
amplitude in different ways. For example, subprocesses with
definite instants of the beginning and  end of the interaction in
the system can be considered as such alternatives. In this way the
amplitude $<\psi_2|U(t,t_0)|\psi_1>$  can be written in the form
[3]
\begin{eqnarray}
<\psi_2| U(t,t_0)|\psi_1> = <\psi_2|\psi_1> +\nonumber\\
+\int_{t_0}^t dt_2 \int_{t_0}^{t_2} dt_1 <\psi_2|\tilde
S(t_2,t_1)|\psi_1>, \label{repre}
\end{eqnarray}
where $<\psi_2|\tilde S(t_2,t_1)|\psi_1>$ is the probability
amplitude that if at time $t_1$ the system was in the state
$|\psi_1>,$ then the interaction in the system will begin at time
$t_1$ and will end at  time $t_2,$ and at this time the system
will be in the state $|\psi_2>.$ Note that in general $\tilde
S(t_2,t_1)$  may be only an operator-valued generalized function
of $t_1$ and $t_2$ [3], since only $U(t,t_0)={\bf 1}+
\int^{t}_{t_0} dt_2 \int^{t_2}_{t_0}dt_1\tilde S(t_2,t_1)$  must
be an operator on the Hilbert space. Nevertheless, it is
convenient to call $\tilde S(t_2,t_1)$ an "operator", using this
word in generalized sense. In the case of an isolated system the
operator $\tilde S(t_2,t_1)$ can be represented in the form [3]
\begin{equation}
\tilde S(t_2,t_1) = exp(iH_0t_2) \tilde T(t_2-t_1) exp(-iH_0 t_1),
\label{t}
\end{equation}
$H_0$ being the free Hamiltonian.

As has been shown in Ref.[3], for the evolution operator
$U(t,t_0)$ given by (3) to be unitary for any times $t_0$ and $t$,
the operator $\tilde S(t_2,t_1)$ must satisfy the following
equation:
\begin{eqnarray}
(t_2-t_1) \tilde S(t_2,t_1) = \int^{t_2}_{t_1} dt_4
\int^{t_4}_{t_1}dt_3 \nonumber \\
 \times(t_4-t_3) \tilde S(t_2,t_4) \tilde S(t_3,t_1).
\label{main}
\end{eqnarray}
This equation allows one to obtain the operators $\tilde
S(t_2,t_1)$ for any $t_1$ and $t_2$, if the operators $\tilde
S(t'_2, t'_1)$ corresponding to infinitesimal duration times $\tau
= t'_2 -t'_1$ of interaction are known. It is natural to assume
that most of the contribution to the evolution operator in the
limit $t_2 \to t_1$ comes from the processes associated with the
fundamental interaction in the system under study. Denoting this
contribution by $H_{int}(t_2,t_1)$, we can write
\begin{equation}
\tilde{S}(t_2,t_1) \tend\limits_{t_2\rightarrow t_1}
H_{int}(t_2,t_1) + o(\tau^{\epsilon}), \label{bound}
\end{equation}
where $\tau=t_2-t_1$. The parameter $\varepsilon$ is determined by
demanding that $H_{int}(t_2,t_1)$ must be so close to the solution
of Eq.(5) in the limit $t_2\tend t_1$ that this equation has a
unique solution having the behavior (6) near the point
$t_2=t_1$.Thus this operator must satisfy the condition
\begin{eqnarray}
(t_2-t_1) H_{int}(t_2,t_1)\tend\limits_{t_2 \tend t_1}
\int^{t_2}_{t_1} dt_4 \int^{t_4}_{t_1} dt_3 (t_4-t_3)\nonumber\\
 \times
H_{int}(t_2,t_4) H_{int}(t_3,t_1)+ o(\tau^{\epsilon+1}).
\label{bound'}
\end{eqnarray}
Note that the value of the parameter $\epsilon$ depends on the
form of the operator $ H_{int}(t_2,t_1).$ Since $\tilde
S(t_2,t_1)$ and $H_{int}(t_2,t_1)$ are only operator-valued
distributions, the mathematical meaning of the conditions (6) and
(7) needs to be clarified. We will assume that the condition (6)
means that
\begin{eqnarray}
<\Psi_2|\int^{t}_{t_0} dt_2 \int^{t_2}_{t_0}dt_1\tilde S(t_2,t_1)
|\Psi_1>\tend\limits_{t\tend t_0}\nonumber\\
<\Psi_2|\int^{t}_{t_0} dt_2 \int^{t_2}_{t_0}dt_1
H_{int}(t_2,t_1)|\Psi_1>+o(\tau^{\epsilon+2}), \nonumber
\end{eqnarray}
 for
any vectors $|\Psi_1>$ and $|\Psi_2>$ of the Hilbert space. The
condition (7) has to be  considered in the same sense.

Within the GQD the operator $H_{int}(t_2,t_1)$ plays the role
which the interaction Hamiltonian plays in the ordinary
formulation of quantum theory: It generates the dynamics of a
system. Being a generalization of the interaction Hamiltonian,
this operator is called the generalized interaction operator. If
$H_{int}(t_2,t_1)$ is specified, Eq.(5) allows one to find the
operator $\tilde S(t_2,t_1).$ Formula (3) can then be used to
construct the evolution operator $U(t,t_0)$ and accordingly the
state vector
\begin{eqnarray}
|\psi(t)> = |\psi(t_0)> +  \int_{t_0}^t dt_2 \int_{t_0}^{t_2} dt_1
\tilde S(t_2,t_1) |\psi(t_0)> \label{psi}
\end{eqnarray}
 at any time $t.$ Thus
Eq.(5) can be regarded as an equation of motion for states of a
quantum system. By using (3) and (4), the evolution operator can
be represented in the form
\begin{eqnarray}
<n_2|U(t,t_0)|n_1>= <n_2|n_1>+\frac{i}{2\pi} \int^\infty_{-\infty}
dx\label{evolution}\nonumber\\
 \times \frac
{\exp[-i(z-E_{n_2})t] \exp[i(z-E_{n_1})t_0]}
{(z-E_{n_2})(z-E_{n_1})} \nonumber\\
 \times<n_2|T(z)|n_1>,
\end{eqnarray}
 where $z=x+iy$, $y>0$, and
\begin{equation}
<n_2|T(z)|n_1> = i \int_{0}^{\infty} d\tau \exp(iz\tau)
<n_2|\tilde T(\tau)|n_1>. \label{tt}
\end{equation}
From (9), for the evolution operator in the Schr{\"o}dinger
picture, we get
\begin{equation}
U_s(t,0)=\frac{i}{2\pi}\int^\infty_{-\infty} dx\exp
(-izt)G(z),\label{schr}
\end{equation}
where
\begin{equation}
<n_2|G(z)|n_1>=\frac{<n_2|n_1>}{z-E_{n_1}}+
\frac{<n_2|T(z)|n_1>}{(z-E_{n_2})(z-E_{n_1})}.\label{lventa}
\end{equation}
Eq.(11) is the well-known expression establishing the connection
between the evolution operator and the Green operator $G(z)$, and
can be regarded as the definition of the operator $G(z)$.

The equation of motion (5) is equivalent to the following equation
for the T matrix [3]:
\begin{equation}
\frac{d T(z)}{dz} =- \sum
\limits_{n}\frac{T(z)|n><n|T(z)}{(z-E_n)^2}, \label{dif}
\end{equation}
with the boundary condition
\begin{equation}
<n_2|T(z)|n_1> \tend \limits_{|z| \tend \infty} <n_2|
B(z)|n_1>,\label{dbound}
\end{equation}
where
$$
B(z) = i \int_0^{\infty} d\tau \exp(iz \tau) H^{(s)}_{int}(\tau),
$$
$\beta=1+\epsilon,$ and $$H^{(s)}_{int}(t_2-t_1) =
 \exp(-iH_0t_2)H_{int}(t_2,t_1) \exp(iH_0t_1)$$
 is the generalized interaction
operator in the Schr{\"o}dinger picture. The solution of Eq.(13)
satisfies the equation
\begin{eqnarray}
<n_2|T(z_1)|n_1> - <n_2|T(z_2)|n_1> = \nonumber \\
=(z_2 -z_1)
\sum_n
 \frac {<n_2|T(z_2)|n><n|T(z_1)|n_1>}
{(z_2-E_n)(z_1-E_n)}. \label{difer}
\end{eqnarray}
This equation in turn is equivalent to the following equation for
the Green operator
\begin{equation}
G(z_1)-G(z_2)=(z_2-z_1)G(z_2)G(z_1).\label{green}
\end{equation}
This is the  Hilbert identity, which in the Hamiltonian formalism
follows from the fact that in this case the evolution operator
(11) satisfies the Schr{\"o}dinger equation, and hence the Green
operator is of the formr
\begin{equation}
G(z)=(z-H)^{-1},\label{reso}
\end{equation}
$H$ being the total Hamiltonian.  At the same time, as has been
shown in Ref.[3], Eq.(5) and hence Eqs.(13) and (15) are unique
consequences of the unitarity condition and the representation (3)
expressing the Feynman superposition principle (the assumption
(iii)). It should be noted that the evolution operator constructed
by using the Schr{\"o}dinger equation can be represented in the
form (3) [9]. Being written in terms of the operators $\tilde
S(t_2,t_1)$, Eq.(5) does not contain operators describing
interaction in the system. It is a relation for $\tilde
S(t_2,t_1)$, which are the contributions to the evolution operator
from the processes with defined instants of the beginning and end
of the interaction in the systems. Correspondingly Eqs.(13) and
(15) are relations for the T matrix. A remarkable feature of  (5)
is that it works as a recurrence relation, and to construct the
evolution operator it is enough to now the contributions to this
operator from the processes with infinitesimal duration times of
interaction, i.e. from the processes of a fundamental interaction
in the system. This makes it possible to use (5) as a dynamical
equation. Its form does not depend on the specific feature of the
interaction (the Schr{\"o}dinger equation, for example, contains
the interaction Hamiltonian). Since Eq.(5) must be satisfied in
all the cases, it can be considered as the most general dynamical
equation consistent with the current concepts of quantum theory.
All the needed dynamical information contains in the boundary
condition for this equation, i.e. in the generalized interaction
operator $H_{int}(t_2,t_1)$. As has been shown in Ref.[3], the
dynamics governed by Eq.(5) is equivalent to the Hamiltonian
dynamics in the case where the operator $H_{int}(t_2,t_1)$ is of
the form
\begin{equation}
 H_{int}(t_2,t_1) = - 2i \delta(t_2-t_1)
 H_{I}(t_1) ,
 \label{delta}
\end{equation}
$H_{I}(t_1)$ being the interaction Hamiltonian in the interaction
picture. In this case the state vector $|\psi(t)>$ given by (8)
satisfies the Schr{\"o}dinger equation
$$
  \frac {d |\psi(t)>}{d t} = -iH_I(t)|\psi(t)>.
$$
The delta function $\delta(\tau)$ in (18) emphasizes that in this
case the fundamental interaction is instantaneous. Thus the
Schr{\"o}dinger equation results from the generalized equation of
motion (5) in the case where the interaction generating the
dynamics of a quantum system is instantaneous. At the same time,
Eq.(5) permits the generalization to the case where the
interaction generating the dynamics of a quantum system is
nonlocal in time [3,6]. In the general case, the generalized
interaction operator has the following form [7]:
$$H_{int}(t_2,t_1)=-2i\delta(t_2-t_1)H_I(t_1)+H_{non}(t_2,t_1),$$
where the first term on the right-hand side of this equation
describes the instantaneous part of the interaction generating the
dynamics of a quantum system, while the term $H_{non}(t_2,t_1)$
represents its nonlocal-in-time part. As has been shown [3,7],
there is one-to-one correspondence between nonlocality of
interaction and the UV behavior of the matrix elements of the
evolution operator as a function of momenta of particles: The
interaction operator can be nonlocal in time only in the case
where this behavior is "bad", i.e. in a local theory it results in
UV divergences. In Ref.[9] it has been shown that after
renormalization the dynamics of the three-dimensional theory of a
neutral scalar field interacting through a $\varphi^4$ coupling is
governed by the generalized dynamical equation (5) with a
nonlocal-in-time interaction operator. This gives reason to expect
that after renormalization the dynamics of an EFT is also governed
by this equation with a nonlocal interaction operator. In the next
section we will consider this problem by using a toy model of the
separable NN interaction.

\section{Renormalization and nonlocality of the NN interaction}
Let us consider the evolution problem for two nonrelativistic
particles in the c.m.s. We denote the relative momentum by ${\bf
p}$ and the reduced mass by $\mu.$ Assume that the generalized
interaction operator in the Schr{\"o}dinger picture
$H^{(s)}_{int}(\tau)$ has the form
$$
<{\bf p}_2| H^{(s)}_{int}(\tau) |{\bf p}_1> = \varphi^*({\bf p}_2)
\varphi({\bf p}_1) f(\tau),
$$
where $f(\tau)$ is some function of $\tau,$ and the form factor
$\varphi ({\bf p})$ has the following asymptotic behavior for
$|{\bf p}| \tend \infty:$
\begin{equation}
\varphi({\bf p}) \sim |{\bf p}|^{-\alpha}, \quad {(|{\bf p}| \tend
\infty).} \label{form}
\end{equation}
Let, for example, $\varphi ({\bf p})$ be of the form
$$
\varphi({\bf p}) = |{\bf p}|^{-\alpha}+g({\bf {p}}),
$$
and in the limit $|{\bf p}| \tend \infty$ the function $g({\bf
{p}})$ satisfies the estimate $g({\bf {p}})=o(|{\bf
{p}}|^{-\delta})$, where $\delta>\alpha,$ $\delta>\frac{3}{2}.$ In
the separable case, $<{\bf p}_2| \tilde S(t_2,t_1) |{\bf p}_1>$
can be represented in the form
$$
<{\bf p}_2| \tilde S(t_2,t_1) |{\bf p}_1>=\varphi^*({\bf
p}_2)\varphi({\bf p}_1)\tilde s(t_2,t_1).
$$
Correspondingly,
 $<{\bf p}_2| T(z) |{\bf p}_1>$ is of the form
\begin{equation}
  <{\bf p}_2| T(z)|{\bf p}_1> = \varphi^* ({\bf p}_2)\varphi ({\bf p}_1)
t(z),\label{separ}
\end{equation}
 where, as it follows from (13), the function $t(z)$
satisfies the equation
\begin{equation}
\frac {dt(z)}{dz} = -t^2(z) \int d^3k \frac {|\varphi ({\bf
k})|^2} {(z-E_k)^2} \label{deq}
\end{equation}
with the asymptotic condition
\begin{equation}
t(z)  \tend \limits_{|z| \tend \infty} f_1(z) + o(|z|^{-\beta}),
\label{asym}
\end{equation}
where
\begin{equation}
f_1(z)= i \int_0^{\infty} d\tau exp(iz\tau) f(\tau), \label{fn}
\end{equation}
 and $E_k =
\frac {k^2}{2 \mu}.$ The solution of Eq.(21) with the initial
condition $t(a)=g_a,$ where $a \in (-\infty,0),$ is
\begin{equation}
t(z) = g_a \left(1 +(z-a) g_a
 \int d^3k \frac {|\varphi ({\bf k})|^2}
{(z-E_k)(a-E_k)} \right)^{-1}.\label{deqa}
\end{equation}
In the case $\alpha >\frac{1}{2}$, the function $t(z)$ tends to a
constant as $|z| \tend \infty$:
\begin{equation}
t(z)  \tend \limits_{|z| \tend \infty} \lambda.\label{lambda}
\end{equation}
Thus in this case the function $f_1(z)$ must also tend to
$\lambda$ as $|z| \tend \infty.$ From this it follows that the
only possible form of the function $f(\tau)$ is
$$
f(\tau) = -2i \lambda \delta(\tau) + f^{\prime}(\tau),
$$
where the function $f^{\prime}(\tau)$ has no such a singularity at
the point $\tau=0$ as the delta function. In this case  the
generalized interaction operator $H^{(s)}_{int}(\tau)$ has the
form
\begin{equation}
<{\bf p}_2| H^{(s)}_{int}(\tau) |{\bf p}_1>=-2i \lambda
\delta(\tau)\varphi^*({\bf p}_2) \varphi({\bf
p}_1),\label{instant}
\end{equation}
 and hence the dynamics generated by this operator is
equivalent to the dynamics governed by the Schr{\"o}dinger
equation with the separable potential
\begin{equation}
<{\bf p}_2|H_I|{\bf p}_1> = \lambda \varphi^*({\bf p}_2)
\varphi({\bf p}_1).\label{ham}
\end{equation}
Solving Eq.(21) with the boundary condition (25), we easily get
the well-known expression for the T matrix in the
separable-potential model
\begin{equation}
<{\bf p}_2|T(z)|{\bf p}_1> = \lambda \varphi^* ({\bf
p}_2)\varphi({\bf p}_1) \left (1 - \lambda \int d^3k \frac
{|\varphi({\bf k})|^2}{z-E_k} \right )^{-1}.\label{sol}
\end{equation}

Ordinary quantum mechanics does not permit the extension of the
above model to the case $\alpha \leq \frac{1}{2}.$ Indeed, in the
case of such a large-momentum behavior of the form factors
$\varphi({\bf p}),$ the use of the interaction Hamiltonian given
by (28) leads to the UV divergences, i.e. the integral in (29) is
not convergent. We will now show that the generalized dynamical
equation (5) allows one to extend this model to the case
$-\frac{1}{2} < \alpha <\frac{1}{2}.$ Let us determine the class
of the functions $f_1(z)$ and correspondingly the value of $\beta$
for which Eq.(21) has a unique solution having the asymptotic
behavior (22). In the case $\alpha <\frac{1}{2},$ the function
$t(z)$ given by (24) has the following behavior for $|z| \tend
\infty:$
\begin{equation}
t(z)  \tend \limits_{|z| \tend \infty}  b_1
(-z)^{\alpha-\frac{1}{2}}+ b_2 (-z)^{2 \alpha-1} + o(|z|^{2
\alpha-1}),\label{zero}
\end{equation}
where $b_1 =- \frac{1}{2} cos(\alpha \pi) \pi^{-2} (2
\mu)^{\alpha-\frac{3}{2}}$ and $b_2= b_1 |a|^{\frac{1}{2}- \alpha}
-b_1^2(M(a)+g_a^{-1})$ with
$$
M(a) = \int \frac {|\varphi({\bf k})|^2-
 |{\bf {k}}|^{-2\alpha}}
{a-E_k}d^3k .
$$
The parameter $b_1$ does not depend on $g_a.$ This means that all
solutions of Eq.(21) have the same leading term  in (29), and only
the second term distinguishes the different solutions of this
equation. Thus, in order to obtain a unique solution of Eq.(21),
we must specify the first two terms in the asymptotic behavior of
$t(z)$ for $|z| \tend  \infty.$ From this it follows that the
functions $f_1(z)$ must be of the form
$$
f_1(z) = b_1 (-z)^{\alpha-\frac{1}{2}} + b_2 (-z)^{2 \alpha -1} ,
$$
and $\beta=2 \alpha -1.$ Correspondingly, the functions $f(\tau)$
must be of the form
\begin{equation}
f(\tau) = a_1 \tau^{-\alpha-\frac{1}{2}} + a_2 \tau^{-2 \alpha},
\end{equation}
with $a_1= -ib_1 \Gamma ^{-1}(1-2\alpha) exp[i(-\frac{\alpha}{2}+
\frac{1}{4}) \pi],$ and $a_2= -b_2 \Gamma ^{-1}(1-2\alpha) exp(-i
\alpha \pi),$ where $\Gamma(z)$ is the gamma-function. This means
that in the case $\alpha <\frac{1}{2}$ the generalized interaction
operator must be of the form
\begin{eqnarray}
<{\bf p}_2| {H}^{(s)}_{int}(\tau)|{\bf p}_1>
 = \varphi^* ({\bf
p}_2)\varphi ({\bf p}_1)\nonumber\\
\times\left( a_1\tau^{-\alpha-\frac{1}{2}} + a_2 \tau^{-2
\alpha}\right), \label{less}
\end{eqnarray}
and, as it follows from (20) and (24), for the T matrix we have
\begin{equation}
<{\bf p}_2| T(z)|{\bf p}_1> = N(z) \varphi^* ({\bf p}_2)\varphi
({\bf p}_1),\label{nz}
\end{equation}
with
$$
N(z) = g_a \left (1 + (z-a) g_a \int d^3 k \frac{|\varphi({\bf
k})|^2} {(z-E_k)(a- E_k)} \right )^{-1},
$$
where $$ g_a = b_1^2\left (b_1 |a|^{\frac{1}{2} -\alpha} + a_2
\Gamma(1-2\alpha) exp( i \alpha \pi)-b_1^2M(a)\right )^{-1}. $$
It
can be easily checked that $N(z)$ can be represented in the
following form
$$
N(z)=\frac{b_1^2}{-b_2+b_1(-z)^{\frac{1}{2}-\alpha}+M(z)b_1^2}.
$$

By using (9) and (32), we can construct the evolution operator
\begin{eqnarray}
<{\bf p}_2|U(t,t_0)|{\bf p}_1> = <{\bf p}_2|{\bf p}_1> + \frac
{i}{2\pi} \int_{-\infty}^{\infty} dx \nonumber\\
\times \frac {\exp[-i(z-E_{p_2})t]
\exp[i (z - E_{p_1})t_0]} {(z-E_{p_2})(z-E_{p_1})} \nonumber\\
\times \varphi^*({\bf p}_2) \varphi({\bf p}_1)N(z), \label{evol}
\end{eqnarray}
 where
$z=x+iy,$ and $y>0.$ The evolution operator $U(t,t_0)$ defined by
(33) is a unitary operator satisfying the composition law (2),
provided that the parameter $b_2$ is real.

Let us now consider the case $\alpha =\frac{1}{2}$. In this case
the UV behavior of the form factor $\varphi({\bf p})$ gives rise
to the logarithmic singularities in the Born series. For $\alpha
=\frac{1}{2}$ the function $t(z)$ given by (33) has the following
asymptotic behavior:
$$
t(z)  \tend \limits_{|z| \tend \infty}  b_1 \ln^{-1}(-z)+ b_2
\ln^{-2}(-z) + o(\ln^{-2}(-z)),
$$
where $b_1 =- (4\pi \mu)^{-1}$ and $b_2= b_1 \ln(-a)
-b_1^2(M(a)+g_a^{-1})$.  From this it follows that the functions
$f_1(z)$ must be of the form
\begin{equation}
f_1(z) = b_1\ln^{-1}(-z)+ b_2 \ln^{-2}(-z), \label{ln}
\end{equation}
and $\beta=2 \alpha -1.$ By using the fact that, as it follows
from (23) and (34),
$$
f(\tau)= -\frac{i}{2\pi}
 \int_{-\infty}^{\infty}dx
 \exp(-iz\tau)\left(\frac{b_1}{\ln(-z)}+\frac{b_2}{
 \ln^{2}(-z)}\right)
$$
we can obtain the interaction operator
\begin{eqnarray}
 <{\bf p}_2| {H}^{(s)}_{int}(\tau)|{\bf p}_1> = -\frac{i}{2\pi}
\varphi^* ({\bf p}_2) \varphi ({\bf
 p}_1)\nonumber \\
 \times\int_{-\infty}^{\infty}dx
 \exp(-iz\tau)\left(\frac{b_1}{\ln(-z)}+\frac{b_2}{
 \ln^{2}(-z)}\right).\label{equal}
\end{eqnarray}
 Thus, in the case of the UV behavior of the form factors
corresponding to $\alpha=\frac{1}{2}$, the interaction operator
$H^{(s)}_{int}(\tau)$ necessarily has the form (35), where for
given $\psi({\bf p})$ only the parameter $b_2$ is free. The
solution of Eq.(13) with this interaction operator is of the form
(32), where
$$
 N(z)=b_1^2\left(-b_2+b_1\ln(-z)+M(z)b_1^2\right)^{-1}.
$$
\begin{figure}
\resizebox{0.9\columnwidth}{!}{\includegraphics{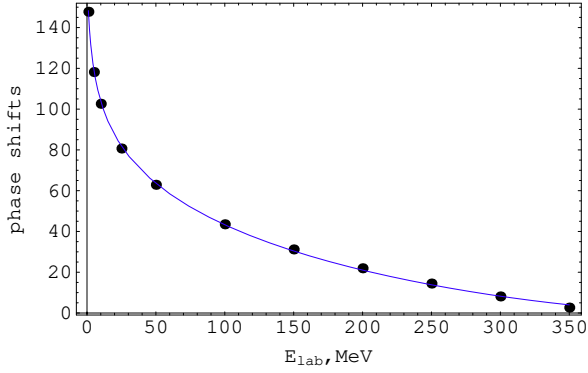}}
\caption{Phase shifts (solid line) in the ${}^3S_1$ channel for np
scattering, compared to the experimental data (points) [8].}
\end{figure}
\begin{figure}
\resizebox{0.9\columnwidth}{!}{\includegraphics{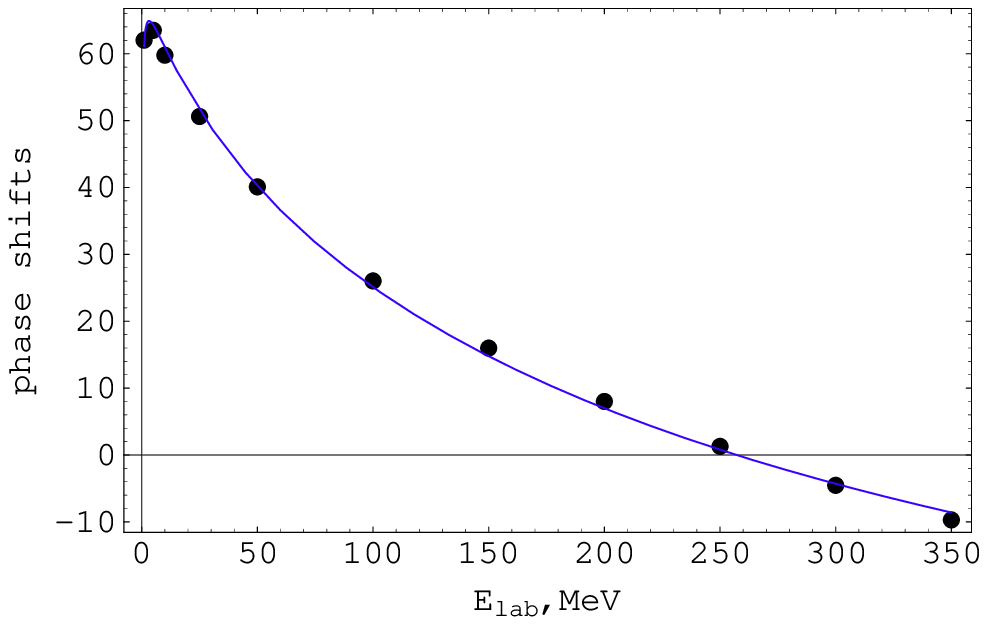}}
\caption{Phase shifts (solid line) in the ${}^1S_0$ channel for np
scattering, compared to the experimental data (points) [8].}
\end{figure}
\begin{figure}
\resizebox{0.9\columnwidth}{!}{\includegraphics{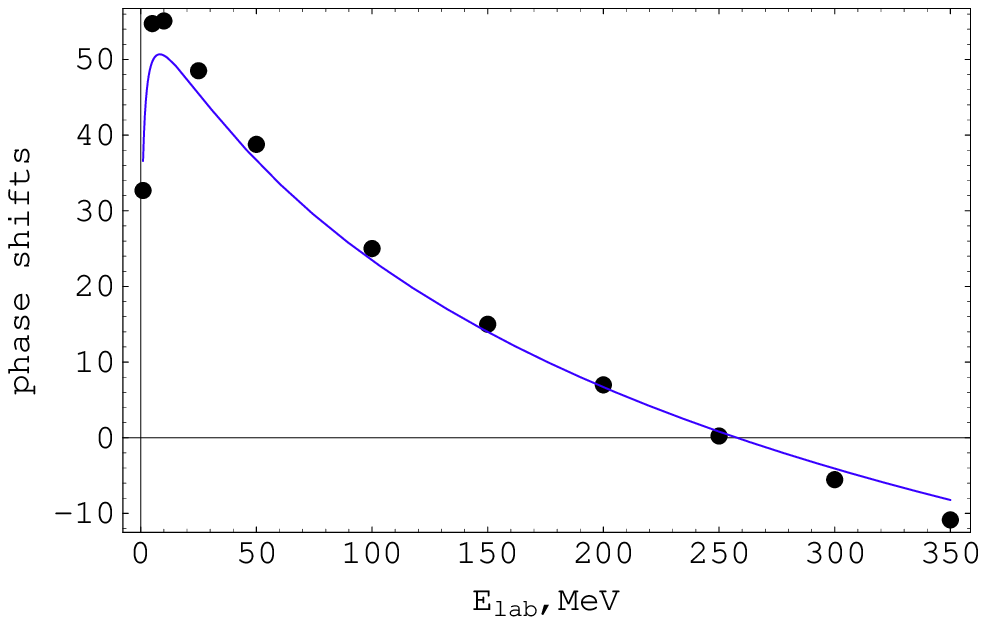}}
\caption{Phase shifts (solid line) in the ${}^1S_0$ channel for pp
scattering, compared to the experimental data (points) [8].}
\end{figure}
\begin{table}
\caption{The parameters of the interaction operator obtained  by
fitting the NN date,  $\rho=1MeV^{-1}$.}
\begin{tabular}{|c|c|c|c|c|c|}
\tableline partial wave&  $\alpha$ &$c_1$ & $\beta\cdot \rho$ &
$d\cdot \rho$ & $ b_2\cdot \rho^{1-2\alpha}$ \\
\tableline ${}^3S_1(np)$ & 0.499 & $133.5\cdot 10^2$ &433.8 &
$766.2$
& $1.696\cdot 10^{-7}$  \\
\tableline ${}^1S_0(np)$ & 0.499 & 131.8 & 356.3 & $3.651\cdot
10^6$  & $1.694\cdot 10^{-7}$ \\
\tableline ${}^1S_0(pp)$ & 0.499 & 320.0 & 371.7 & $6.763\cdot
10^5$  & $1.695\cdot 10^{-7}$ \\
\tableline
\end{tabular}
\end{table}
In Ref.[7] the model for $\alpha<\frac{1}{2}$ was used for
describing the NN interaction at low energies. The motivation to
use such a model for parameterization of the NN forces is the fact
that due to the quark and gluon degrees of freedom the NN
interaction must be nonlocal in time. However, because of the
separation of scales the system of hadrons should be regarded as a
closed system, i.e. the evolution operator must be unitary and
satisfy the composition law. From this it follows that the
dynamics of such a system is governed by Eq.(5) with
nonlocal-in-time interaction. In Ref.[7] the following form factor
was used for  parameterization of the NN interaction
$$
\varphi({\bf p})=\chi({\bf p})+c_1 g_Y({\bf p}), $$ with
$$\chi({\bf p})=\left(d^2+ p^2\right)^{-\frac{\alpha}{2}},
\quad -\frac{1}{2}<\alpha<\frac{1}{2},$$ $g_Y({\bf p})$ being the
Yamaguchi form factor
$$
g_Y({\bf p})=\frac{1}{\beta^2+p^2}. $$ Here $d,$ $c_1$ and $\beta$
are some constants. Being the generalization of the Yamaguchi
model [10] to the case where the NN interaction is nonlocal in
time, our model yields the nucleon-nucleon phase shifts in good
agreement with experiment (see Figs.(1-3)). However, the main
advantage of this model, is that it allows one to investigate the
effects of the retardation in the NN interaction caused by the
existence of the quark and gluon degrees of freedom on nucleon
dynamics.

As we have seen, there is the one-to-one correspondence between
the form of the generalized interaction operator and the UV
behavior of the form factor $\varphi({\bf p}).$ In the case
$\alpha
>\frac{1}{2},$ the operator $H^{(s)}_{int}(\tau)$ would
necessarily have the form (26). In this case the fundamental
interaction is instantaneous. In the case $-\frac{1}{2} <\alpha
<\frac{1}{2}$ (the restriction $\alpha
>-\frac{1}{2}$ is necessary for the integral in (24) to be
convergent), the only possible form of $H^{(s)}_{int}(\tau)$ is
(31), and, in the case $\alpha=\frac{1}{2}$, it must be of the
form (35), and hence the interaction generating  the dynamics of
the system is nonlocal in time. Thus the interaction generating
the dynamics can be nonlocal in time only if the form factors have
the "bad" large-momentum behavior that within Hamiltonian dynamics
gives rise to the ultraviolet divergences:
\begin{mathletters}
\begin{eqnarray}
\text{locality}\quad\Leftrightarrow\varphi({\bf p})\quad\sim
\quad|{\bf
p}|^{-\alpha},\quad \alpha>\frac{1}{2} \nonumber\\
\text{nonlocality}\Leftrightarrow\varphi({\bf
p})\quad\sim\quad|{\bf p}|^{-\alpha},\quad
\alpha\leq\frac{1}{2}\nonumber
\end{eqnarray}
\end{mathletters}
From this it follows that the quark-gluon retardation effects must
results in the "bad" UV behavior of the matrix elements of the
evolution operator as a function of momenta of hadrons. Note that
EFT's lead to the same conclusion:Within the EFT approach the
quark and gluon degrees of freedom manifest themselves in the form
of Lagrangians consistent with the symmetries of QCD which gives
rise to the UV divergences. Note also that EFT's are local
theories, despite the existence of the external quark and gluon
degrees of freedom. However, renormalization of EFT's gives rise
to the fact that these theories become nonlocal. Below this will
be shown by using the example of our toy model.

As we have stated, the interaction operator (31) contains only one
free parameter $b_2$. However, if there is a bound state in the
channel under study, then the parameter $b_2$ is completely
determined by demanding that the T matrix has  the pole at the
bound-state energy. For example, in the ${}^3S_1$ channel the T
matrix has a pole at energy $E_B=-2.2246$MeV. This means  that
$\left[t(E_B)\right]^{-1}=0$,  and, by putting $a=E_B$ in Eq.(24),
we get
\begin{equation}
\left [ t(z)\right ]^{-1}=(z-E_B)\int d^3 k\frac{|\varphi({\bf
{k}})|^2}{(z-E_k) (E_B-E_k)}.\label{our}
\end{equation}
In this case $b_2=b_1\ln(-E_B)-b_1^2M(E_B)$. Let us now show that
renormalization of the LS equation with the separable potential
leading to a logarithmic singularities produces the same T matrix.
In [4] the problem of renormalization of the LS equation was
considered by using the example of the separable potential
\begin{equation}
<{\bf p}_2|V|{\bf p}_1>=\lambda\varphi^*({\bf p}_2)\varphi({\bf
p}_1)
\end{equation}
 with $\psi({\bf p})=\left(d^2+p^2\right)^{-\frac{1}{4}}$. The corresponding T
matrix is of the form (20) with
$$
t(z)=\left (\lambda^{-1} -J(z)\right)^{-1}, $$ where the integral
$$ J(z)=\int d^3k\frac{|\varphi({\bf {k}})|^2}{z-E_k}$$ has a
logarithmic divergence. By using the dimensional regularization,
one can get
$$
\left [t_{\varepsilon}(z)\right ]^{-1}=
\lambda_{\varepsilon}^{-1}-J_{\varepsilon}(z),
$$
where $$J_{\varepsilon}(z)=\int d^{3-\varepsilon}k\frac
{|\varphi({\bf {k}})|^2}{z-E_k}.$$ The strength of the potential
$\lambda_{\varepsilon}$ is adjusted to give the correct
bound-state energy:
$\lambda_{\varepsilon}^{-1}=J_{\varepsilon}(E_B).$ In this way we
get
\begin{eqnarray}
[t_{\varepsilon}(z)]^{-1}=J_{\varepsilon}(E_B)-J_{\varepsilon}(z)
=\nonumber\\
= (z-E_B)\int d^{3-\varepsilon}k\frac {|\varphi({\bf
{k}})|^2}{(z-E_k)(E_B-E_k)}.\label{tEb}
\end{eqnarray}
The right-hand side of (38) is well-behaved in the limit
$\varepsilon\rightarrow 0$. It is easy to see that, taking this
limit, we get the expression (36) for $t(z)$. Thus renormalization
of the LS equation with the above singular potential leads to the
T matrix we have obtained by solving Eq.(13) with nonlocal-in-time
interaction operator. This T matrix satisfies the generalized
dynamical equation (13), but does not satisfy the LS equation.
Correspondingly the Schr{\"o}dinger equation is not  valid in this
case. The strength of the potential $\lambda_\varepsilon$ tends to
zero as $\varepsilon\rightarrow 0$, and consequently the
renormalized interaction Hamiltonian is also tend to zero. Thus
despite the fact that the dynamics of each theory corresponding to
the dimension ${\cal D}=3-\varepsilon$ is Hamiltonian dynamics for
every $\varepsilon>0$, in the limiting case ${\cal D}=3$ we go
beyond Hamiltonian dynamics. The dynamics of the renormalized
theory is governed by the generalized equation of motion (5) with
nonlocal-in-time interaction operator which in the case of the
model under study is given by (35).

\section{Nonlocality in time of the NN interaction and an
anomalous off-shell behavior of two-nucleon amplitudes.}

As we have shown, renormalization in the model, in which the NN
interaction is described by the singular potential, leads to the
fact that the dynamics of the system is governed by the
generalized dynamical equation with a nonlocal-in-time interaction
operator, and in this way we get the same results we have obtained
by using the model constructed within the GQD (in the case
$\alpha=\frac{1}{2}$). This is not surprising. In fact, Eq.(5) is
a consequence of the most general physical principles, and must,
for example, be satisfied in order that the evolution operator be
unitary, and satisfy the composition law (2). Hence, the T matrix
obtained by using a renormalization procedure must satisfy
Eq.(13), provided this procedure leads to physically acceptable
results, i.e. the theory is renormalizable. On the other hand the
freedom in choosing the form of the interaction operator
determining the boundary condition for Eq.(5) is limited by the
condition (7),and, as we have seen, in the case of the form
factors having the UV behavior (19) with $\alpha=\frac{1}{2}$, the
interaction operator must be of the form (35) where only the
parameter $b_2$ is free. However, in the case of the ${}^3S_1$
channel, this parameter is completely determined by demanding that
the T matrix has a pole at energy $E_B$. Thus we have a unique
interaction operator satisfying the above requirements, the use of
which in the boundary condition (14) for Eq.(13) leads to the same
results as renormalization in the separable-potential model.

The main results of the above analyzes of the dynamical situation
in the model under study is that renormalization gives rice to the
fact that the dynamics of a quantum system is governed by the
generalized dynamical equation (5) with nonlocal-in-time
interaction operator. In Ref.[8] this fact has been shown, by
using the example of the three-dimensional theory of a neutral
scalar field interacting through a $\varphi^4$ coupling. This
gives reason to suppose that such a dynamical situation takes
place in any renormalizable theory, for example, in an EFT. Below
we will consider some general argument leading to this conclusion.

Let $G_\Lambda(z)$ be the Green operator of a renormalizable
theory corresponding to the momentum cutoff interaction
Hamiltonian $H_I^{(\Lambda)}(t)$ with renormalized constants. For
every finite cutoff $\Lambda$, the operator $G_\Lambda(z)$
obviously satisfies the Hilbert identity (16)
\begin{equation}
G_\Lambda(z_1)-G_\Lambda(z_2)=(z_2-z_1)G_\Lambda(z_2)G_\Lambda(z_1).
\label{grl}
\end{equation}
At the same time, the renormalized Green operator $G_{ren}(z)$ is
a limit of the consequence of the operators $G_\Lambda(z)$ for
$\Lambda\to\infty$. Since Eq.(39) is satisfied for every
$\Lambda$, and contains only the operator $G_\Lambda(z)$, the
renormalized Green operator must also satisfy this equation
$$
G_{ren}(z_1)-G_{ren}(z_2)=(z_2-z_1)G_{ren}(z_2)G_{ren}(z_1),
$$
despite the fact that the renormalized Green operator cannot be
represented in the form (17) (in the limit $\Lambda\to\infty$ the
operators $H_I^{(\Lambda)}(t)$ do not converge to some operator
acting on the Hilbert space). Correspondingly the renormalized T
matrix satisfies Eqs.(13) and (15), despite the fact that in this
case the LS and Schr{\"o}dinger equations do not follow from these
equations. Here the advantages of the GQD manifest themselves.
Within the GQD the dynamical equation (5) are derived as
consequences of the most general physical principles, and for
Eq.(15) to be satisfied, the Green operator $G(z)$ need not be
represented in the form (17). In the GQD this operator is defined
by (12), where the T matrix in turn is defined by (11), i.e. is
expressed in terms of the amplitudes $<\psi_2|\tilde
S(t_2,t_1)|\psi_1>$ being the contributions to the evolution
operator from the processes in which the interaction in a quantum
system begins at time $t_1$ and ends at time $t_2$. As we have
shown, this allows one to use Eq.(5) and hence Eq.(13) as
dynamical equations: Only in the case where the interaction
operator is of the form (18), i.e. the dynamics of the system is
Hamiltonian, the operator $G(z)$ can be represented in the form
(17). For every finite cutoff $\Lambda$ the dynamics of the system
is Hamiltonian. At the same time, in the limiting case
$\Lambda\tend\infty$ the dynamics is governed by the generalized
dynamical equation with a nonlocal-in-time interaction operator,
i.e. the dynamics is not Hamiltonian.

In order to clarify the characteristic features of the dynamics
generated by the nonlocal-in-time interaction, let us come back to
our toy model. In the Schr{\"o}dinger picture, the evolution
operator
$$
<{\bf p}_2|V(t)|{\bf p}_1>\equiv<{\bf p}_2|U_s(t,0)|{\bf p}_1>$$
of the model (33) can be rewritten in the form
\begin{eqnarray}
<{\bf p}_2|V(t)|{\bf p}_1> =<{\bf p}_2|{\bf p}_1>
\exp(-iE_{p_2}t)\nonumber\\+ \frac {i}{2\pi}
\int_{-\infty}^{\infty} dx \frac {\exp(-izt) <{\bf p}_2|T(z)|{\bf
p}_1>} {(z-E_{p_2})(z-E_{p_1})},
\end{eqnarray}
where $<{\bf p}_2|T(z)|{\bf p}_1>$ is given by (32). Since this T
matrix satisfies Eqs.(13) and (15), the evolution operator (33) is
unitary, and satisfies the composition law (2). Correspondingly,
the operator $V(t)$ constitute a one-parameter group of unitary
operators, with the group property
\begin{equation}
V(t_1+t_2) = V(t_1) V(t_2) , \quad    V(0)= {\bf 1}.
\end{equation}
Assume that this group has a self-adjoint infinitesimal generator
$H$ which in the Hamiltonian formalism is identified with the
total Hamiltonian. Then for $|\psi>\in{\cal D}(H)$ we have
\begin{equation}
\frac{V(t)|\psi> - |\psi>}{t}\tend\limits_{t \tend 0}-iH|\psi>.
\end{equation}
From this and (41) it follows that
$$H=H_0+H_I,$$
where
\begin{eqnarray}
<{\bf p}_2|H_I|{\bf p}_1>= \frac {i}{2\pi} \int_{-\infty}^{\infty}
dx \exp(-izt) \nonumber\\ \times\frac {z<{\bf p}_2|T(z)|{\bf
p}_1>} {(z-E_{p_2})(z-E_{p_1})}.
\end{eqnarray}
Since $<{\bf p}_2|T(z)|{\bf p}_1>$ is an analytic function of $z$,
and, in the case $\alpha\leq\frac{1}{2}$, tends to zero as
$|z|\tend \infty$, from (44) we get $<{\bf p}_2|H_I|{\bf p}_1>=0$
for any ${\bf p}_2$ and ${\bf p}_1$, and hence $H=H_0$. This means
that, if the infinitesimal generator of the group of the operators
$V(t)$ exists, then it coincides with the free Hamiltonian, and
the evolution operator is of the form $V(t)=\exp(-iH_0t)$. Thus,
since this, obviously, is not true, the group of the operators
$V(t)$ has no infinitesimal generator, and hence the dynamics is
not governed by the Schr{\"o}dinger equation.

It should be also noted that in the case $\alpha\leq\frac{1}{2}$,
$\tilde S(t_2,t_1)$ is not an operator on the Hilbert space. In
fact, the function
\begin{eqnarray}
\psi({\bf p})=<{\bf p}|\tilde S(t_2,t_1)|\psi_1>=\nonumber \\
=\varphi^*({\bf p})\tilde S(t_2,t_1)\int d^3k \varphi({\bf
k})<{\bf k}|\psi_1>
\end{eqnarray}
is not square integrable for any nonzero $|\psi_1>$, because of
the slow rate of decay of the form factor $\varphi({\bf p})$ as
$|{\bf p}|\tend\infty$.  Correspondingly, the T matrix given by
(32) does not represent an operator on the Hilbert space. However,
as we have stated, in general $\tilde S(t_2,t_1)$ may be only an
operator-valued generalized function such that the evolution
operator is an operator on the Hilbert space. Correspondingly, the
T matrix must be such that $G(z)$ given by (12) is an operator on
the Hilbert space. In our model, $\tilde S(t_2,t_1)$ and the T
matrix satisfy these requirements, since the evolution operator
given by (33) and the corresponding $G(z)$ are operators on the
Hilbert space. At the same time, in this case we go beyond
Hamiltonian dynamics.

We have shown that after renormalization nucleon dynamics is
governed by the generalized dynamical equation (5) with a
nonlocal-in-time interaction operator. The analyzes of the
dynamical situation arising due to the existence of the external
quark and gluon degrees of freedom leads to the same conclusion.
As we have shown by using our toy model, such a nonlocality of the
NN interaction can have significant effects on the character of
nucleon dynamics. As is well known, the dynamics of many nucleon
systems depends on the off-shell properties of the two-nucleon
amplitudes. For this reason, let us consider the effects of the
nonlocality of the NN interaction on these properties.

 In the nonlocal case, the matrix
elements of the evolution operator as functions of momenta do not
go to zero at infinity so fast as it is required by ordinary
quantum mechanics, and within the Hamiltonian formalism this leads
to the ultraviolet divergences. For example, in this case the
two-nucleon amplitudes $<{\bf p}_2|T(z)|{\bf p}_1>$  do not go to
zero fast enough to make the Faddeev equation well-behaved. Note
that the same problem arises in the EFT approach.
 Another consequence of nonlocality in time of the $NN$ interaction is that
for fixed momenta ${\bf p}_1$ and ${\bf p}_2$ the matrix elements
$<{\bf p}_2|T(z)|{\bf p}_1>$ tend to zero as $|z|\to \infty$,
while, in the local case, they tend to $<{\bf p}_2|V|{\bf p}_1>$
in this limit. To illustrate this, we present in Fig.4 the
off-shell behavior of $<{\bf p}_2|T(z)|{\bf p}_1>$ in the limit
$|z|\to \infty$. Thus, nonlocality in time of the $NN$ interaction
caused by the existence of the quark and gluon degrees of freedom
gives rise to an anomalous off-shell behavior of the two-nucleon
amplitudes. The off-shell properties of the amplitudes for the
ordinary interaction operator and the operator containing the
nonlocal term are qualitatively different. This is true even when
the two interaction operators have approximately the same phase
shifts. Such a large variation in the off-shell behavior of the
amplitudes, even when the interaction operators are identical
on-shell, can have significant effects on three- and many-body
results [10]. This gives reason to expect that the anomalous
off-shell behavior of the two-nucleon amplitudes can also have
significant effects on nucleon matter properties.
\begin{figure}
\resizebox{0.9\columnwidth}{!} {\includegraphics{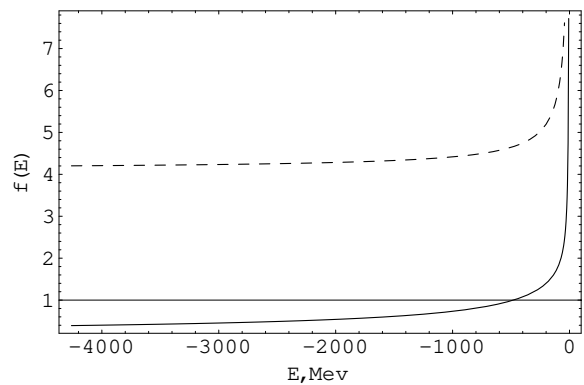}}
\caption{The off-shell behavior of $f(E)=<{\bf p}|T(z)|{\bf p}>,$
$|{\bf p}|=500$ MeV in the ${}^3S_1$ channel for np scattering.
The solid curves corresponds to the model with generalized
interaction operator (31), compared to the model with Yamaguchi
potential with parameters given in [10] (dashed line).}
\end{figure}

\section{Conclusion.}

By using the model of the separable NN potential which gives rise
to logarithmic singularities in the Born series, we have
demonstrated that after renormalization the dynamics of a nucleon
system is governed by the generalized dynamical equation (5) with
a nonlocal in time interaction operator. By using Eq.(16) we have
shown that this should be true in  the general case. At the same
time, being very simple and exactly solvable, the toy model allows
one to investigate some characteristic features of the dynamics of
a theory after renormalization. We have shown that the dynamical
situation in the toy model which arises after renormalization is
completely unsatisfactory from the point of view of the
Hamiltonian formalism: The group of the evolution operators $V(t)$
has no infinitesimal generator, and hence the Hamiltonian of the
renormalized theory cannot be defined. On the other hand, as has
been shown in Ref.[3] , the current concepts of quantum theory
allow the extension of quantum dynamics to the case where the
group of the evolution operator $V(t)$ has no infinitesimal
generator (in this case the interaction generating the dynamics of
a system is nonlocal in time). The Schr{\"o}dinger equation is
only a particular case of the generalized dynamical equation (5)
derived as a consequence of the most general principles of quantum
theory, and there are no reasons to restrict ourselves to the
local case where the interaction is instantaneous and the
evolution operator has an infinitesimal generator. Moreover, in
quantum field theory such a restriction gives rise to the UV
divergences, and, as has been shown, after renormalization the
dynamics of a theory is not Hamiltonian. At the same time, we have
shown that such a dynamics is governed by the generalized
dynamical equation (5) with a nonlocal-in-time interaction
operator. This means that the GQD provides the extension of
quantum dynamics which is needed for describing the evolution of
quantum systems within a renormalized theory. This gives reason to
hope that the use of the GQD and parameterization of the NN forces
like (35) can open new possibilities for applying the EFT approach
to the description of low-energy nucleon dynamics. By using an
EFT, one can construct the generalized interaction operator
consistent  with the symmetries of QCD. This operator can then be
used in Eq.(5) for describing nucleon dynamics. Note that in this
case regularization and renormalization are needed only on the
stage of determining the generalized interaction operator. After
this one does not face the problem of UV divergences.


\end{document}